\documentclass[12pt,aasmacros]{spieman}  
\usepackage{amsmath,amsfonts,amssymb}
\usepackage{graphicx}
\usepackage{setspace}
\usepackage{tocloft}
\usepackage{xcolor}
\usepackage{xspace}
\usepackage{lineno}

\newcommand{\NuSTAR}{{\it NuSTAR}\xspace}
\newcommand{\ProtoEXISTa}{{\it ProtoEXIST1}\xspace}

\newcommand{\HREXI}{{HREXI}\xspace}

\newcommand{\um}{\mbox{$\mu$m}\xspace}

\newcommand{\fixR}[1]{{#1}}
\newcommand{\fixU}[1]{{#1}}

\title{Proof of Concept for Through Silicon Vias (TSVs) in Application Specific Integrated Circuits (ASICs) for Hard X-ray \fixU{Imaging} Detectors}

\author[1]{Jaesub Hong*}
\author[1]{Jonathan Grindlay}
\author[1]{Branden Allen}
\author[1]{Daniel P. Violette}
\author[2]{Hiromasa Miyasaka}
\author[3]{Dean Malta}
\author[3]{Jennifer Ovental}
\author[3]{David Bordelon}
\author[3]{Daniel Richter}

\affil[1]{Center for Astrophysics $|$ Harvard \& Smithsonian, 60 Garden St, Cambridge, MA 02138, USA}
\affil[2]{California Institute of Technology, Pasadena, CA 91125, USA}
\affil[3]{Micross Advanced Interconnect Technology, Research Triangle Park, NC 27709, USA}

\cftpagenumbersoff{figure}
\cftpagenumbersoff{table} 
\begin{document} 
\maketitle

\begin{abstract}
Application Specific Integrated Circuits (ASICs) are commonly used to efficiently process the signals from sensors and detectors in space. Wire bonding is a space qualified technique of making interconnections between ASICs and their substrate packaging board for power, control and readout of the ASICs. Wire bonding is nearly ubiquitous in modern space programs, but their exposed wires can be prone to damage during assembly and subject to electric interference during operations. 
Additional space around the ASICs needed for wire bonding also impedes efficient packaging of large arrays of detectors. Here we introduce the Through Silicon Vias (TSV) technology that replaces wire bonds and eliminates their shortcomings. We have successfully demonstrated the feasibility of implementing TSVs to existing ASIC wafers (a.k.a. a via-last process) developed for processing the X-ray signals from the X-ray imaging CdZnTe detectors on the Nuclear Spectroscopic Telescope Array (\NuSTAR) Small Explorer telescope mission that was launched in 2012. While TSVs are common in the semiconductor industry, this is the first (to our knowledge) successful application for Astrophysics imaging instrumentation.  We expect that the TSV technology will simplify the detector assembly, and thus will enable significant cost and schedule savings in assembly of large area CdZnTe detectors.
\end{abstract}



{\noindent \footnotesize\textbf{*}Jaesub Hong,  \linkable{jhong@cfa.harvard.edu} }


\section{Introduction \& Motivation}


\fixU{The discovery and study of the most energetic transient astrophysics phenomena (in order of decreasing luminosity) include Gamma-ray bursts of both long and short duration, outbursts from supermassive black holes (particularly blazars) in galactic nuclei, black hole and neutron star mergers discovered as gravitational wave outbursts, flaring outbursts from black hole and neutron star X-ray binaries and extreme flares from single M-dwarf stars. These studies are best pursued by wide-field coded aperture hard X-ray telescopes. 
The High Resolution Energetic X-ray Imager (\HREXI) is the {\it high resolution imaging 3 - 300 keV X-ray detector} designed for optimum readout of a wide-field coded aperture telescope.  Two wide-field, high resolution mission concepts that employ the \HREXI architecture are described in Grindlay et al.~(2021).\cite{Grindlay21}}

\fixU{\HREXI is composed of close-tiled high spatial resolution CdZnTe (CZT) detectors.} The main heritage of CZT detectors in \HREXI  comes from the Application Specific Integrated Circuit (ASIC) used in {Nuclear Spectroscopic Telescope Array} (\NuSTAR) \cite{Harrison13} and the large CZT detector plane (256 cm$^2$) assembled for a series of balloon-borne wide-field hard X-ray imaging telescope experiments \ProtoEXISTa \& {\it 2}. \cite{Hong09,Allen11,Hong13,Hong17}
\fixU{Each 0.3~cm thick CZT crystal in \HREXI has an active imaging area of 1.92~$\times$~1.92~cm$^2$ with 32~$\times$~32 0.6~mm pixels that are each conductive-epoxy bonded to the corresponding 32~$\times$~32 pixels on a \NuSTAR ASIC (hereafer NuASIC).  \HREXI groups CZT-NuASICs into a 2 x 2 close-tiled Detector Crystal Array (DCA) that is read out and individually controlled as a single unit. But unlike \NuSTAR, \HREXI allows DCAs themselves to be close-tiled into arbitrarily large area detector arrays for large area  coded aperture imaging with the wide-field and sensitivity needed for the discovery and study of the energetic transients listed above.} What is needed to make \fixU{\HREXI-based} missions feasible is a technology to enable high yield fabrication and assembly of a large area array of high resolution CZT detectors \fixU{for multiple wide-field telescopes at relatively} low cost. 

For \ProtoEXISTa \& {\it 2} and \NuSTAR, the electrical connection between the NuASIC and the substrate board was made through wire bonding, where each of the 87 wire-bond pads on the NuASIC is connected 
to the matching pad on the substrate board underneath through a thin metal wire.
Wire bonding is a space-qualified technique, commonly used for many electronics devices including detectors and sensors.
However, the exposed wires are prone to damage during handling, and they can complicate the assembly procedure for a large detector array. They can be a source of electrical noise pickup during operation. Protecting wire bonds through potting or similar methods can also invite additional electronics noise. 
In addition, their presence introduces a significant gap\footnote{The NuASIC extends about 2.5 mm from the edge pixels on the wire-bond pad side, and the matching pads on the substrate board requires an additional 2.5 mm gap for safe wire bond installation and protection during assembly. Thus, with wire bonds, the gap between CZT crystals is about 5 mm on the wire-bond side. With TSVs, this gap can shrink down to 2.5 mm for the current NuASICs, and the layout of new ASICs in future can be designed to further minimize this gap with TSVs (see Section~\ref{s:future}).} between CZT detectors, making it difficult to closely tile the detectors into a large array. The large gaps between the detectors 
introduce additional background
X-ray events and spatial non-uniformity in the background by allowing X-rays to enter the detectors through the side walls of CZT crystals (e.g., Figure 7 in Hong et al. 2013).\cite{Hong13}

For the last few years, we have been pursuing the Through Silicon Via (TSV) technology to replace  wire bonds and thus eliminate 
their shortcomings in CZT detector packaging. With TSVs, the power, control and readout lines in an ASIC can be directly connected to the substrate board through
the \fixU{full} Si layer in the ASIC, enabling flip-chip style bonding. Therefore, TSVs make the subsequent assembly procedures of detectors robust and simple, and  it is projected that they would enable significant saving in cost and schedule for the assembly of a large array CZT detector plane. 
TSVs are implemented at a wafer level on multiple wafers in a single run, 
whereas wire bonding requires labor on the individual ASIC dies.
CZT detectors can be tiled more tightly with TSVs by minimizing wasted space in between detectors.
In this paper, we present the first successful implementation of TSVs in the current \NuSTAR wafers.

An increasing number of vendors in the semiconductor industry now routinely utilize TSVs for `vertical' (3-D) packaging of ICs such as 
memory modules. The typical size of these TSVs is $\sim$ 5 \um in diameter 
and 15$-$50 \um in depth. They are often designed and fabricated in parallel with the IC circuitry under the careful layout of stacking geometry: 
i.e., they are implemented through a `via-middle' process, where TSVs are inserted during foundry runs for the ICs, typically after transistor fabrication but prior to back end of the line (BEOL) metalization.

While the via-middle procedure is the most safe approach, redesigning existing ASICs for TSVs requires considerable resources, often prohibitively expensive under the relatively low funding of astrophysics instrumentation programs.  Instead, we have decided to first demonstrate the feasibility and benefits of TSVs through a `via-last procedure', where the TSVs are inserted into the pre-fabricated ASICs on a wafer. The 'via-last' approach can also be applied to other similar programs that can benefit from removal of wire bonds.
One can further divide the via-last procedure into two different approaches: one inserting {\it blind}-TSVs from the back side of the wafer, and the other inserting {\it through}-TSVs from the front side. 
The successful TSV implementation demonstrated in this paper has utilized both approaches. 

\fixU{In this paper, w}e describe the methodology and performance of the back-side {\it blind}-TSV approach by Vendor \#1 in Sections \ref{s:implement} and \ref{s:performance}. 
In Section \ref{s:frontside}, we describe the concept and advantages of the front-side {\it through}-TSV insertion by Vendor \#2 - Micross Advanced Interconnect Technology (AIT)\footnote{http://www.micross.com}, and present the results of the performance measurements from the initial test run.
Section \ref{s:future} presents our vision of future detector assembly and packaging concept using the TSV technology.


\section{Implementing Back-side \textbf{\textit{blind}}-TSVs on NuASIC Wafers} \label{s:implement}

\begin{figure*}[tbh!]
    \includegraphics[width=6.6in]{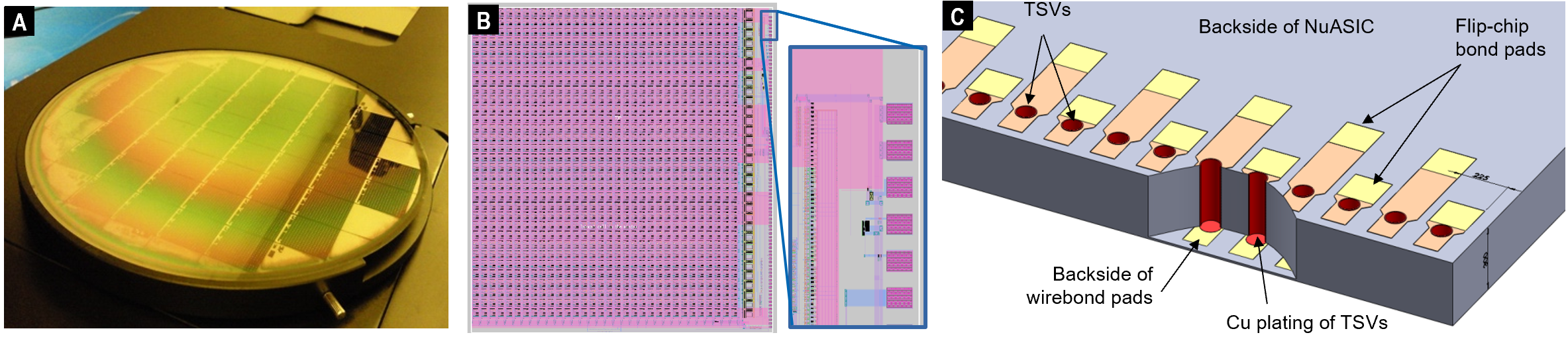}
\caption{(A) A mechanical wafer with 49 ASIC dies was used to develop the TSV process. (B) Each \NuSTAR ASIC has 87 wire-bond pads on one edge for power, control and readout. The inset shows a closeup view of 6 wire-bond pads with 225 $\mu$m pitch. (C) Cutaway view of back-side {\it blind}-TSV concept for \NuSTAR ASIC (NuASIC). TSVs connect the flip-chip bonding pads on the back side of the ASIC to the back of the wire-bond pads on the front side of the NuASIC.}
    \label{f:backsideTSVconcept}
\end{figure*}

NuASICs are fabricated \fixU{by ON Semiconductor Corp.\footnote{www.onsemi.com}}~on 700 \um thick 8 inch silicon wafers. Each wafer has 49 NuASIC dies. 
Figure \ref{f:backsideTSVconcept}A shows a picture of a mechanical (dummy) version of the \NuSTAR wafer, which was used for the process development of TSV insertion. Each NuASIC is equipped with 1024 \fixU{gold stud pixels} in a 32 $\times$ 32 2-D array, matching the anode pattern of 1024 \fixU{conductive epoxy dot} pixels \fixU{o}n a CZT crystal (0.6 mm pixel pitch), 
to which the NuASIC bonds directly \cite{Harrison10, Hong13}.
The 87 wire-bond pads, with 225 $\mu$m pitch, that require 
electric connection are located along one edge of \fixU{each NuASIC die} (the right hand side in Figure \ref{f:backsideTSVconcept}B). The silicon substrate underneath the wi\fixU{r}ebond pads is free of circuitry, making the \NuSTAR wafers suitable for the via-last procedure.

 Figure \ref{f:backsideTSVconcept}C illustrates a conceptual cutaway view of the back-side {\it blind}-TSVs that connect the flip-chip style bonding pads on the back side of the ASIC to the aluminum wire-bond pads on the front side. The basic idea for the back-side TSVs is to implement about 100 \um diameter by 300 \um deep TSVs from the back side of the wafer and touch down on the back of the wire-bond pads. The NuASIC has three metal layers and the wire-bond pads have all three layers electrically connected through internal vias to maximize the mechanical stability during the wire-bonding process.  Thus, connecting the TSVs to the back of the metal 3 layer is equivalent to the electric connection to the wire-bond pad on the top metal layer. 

\begin{figure*}[tbh!]
   \centering
   \includegraphics[width=4.5in]{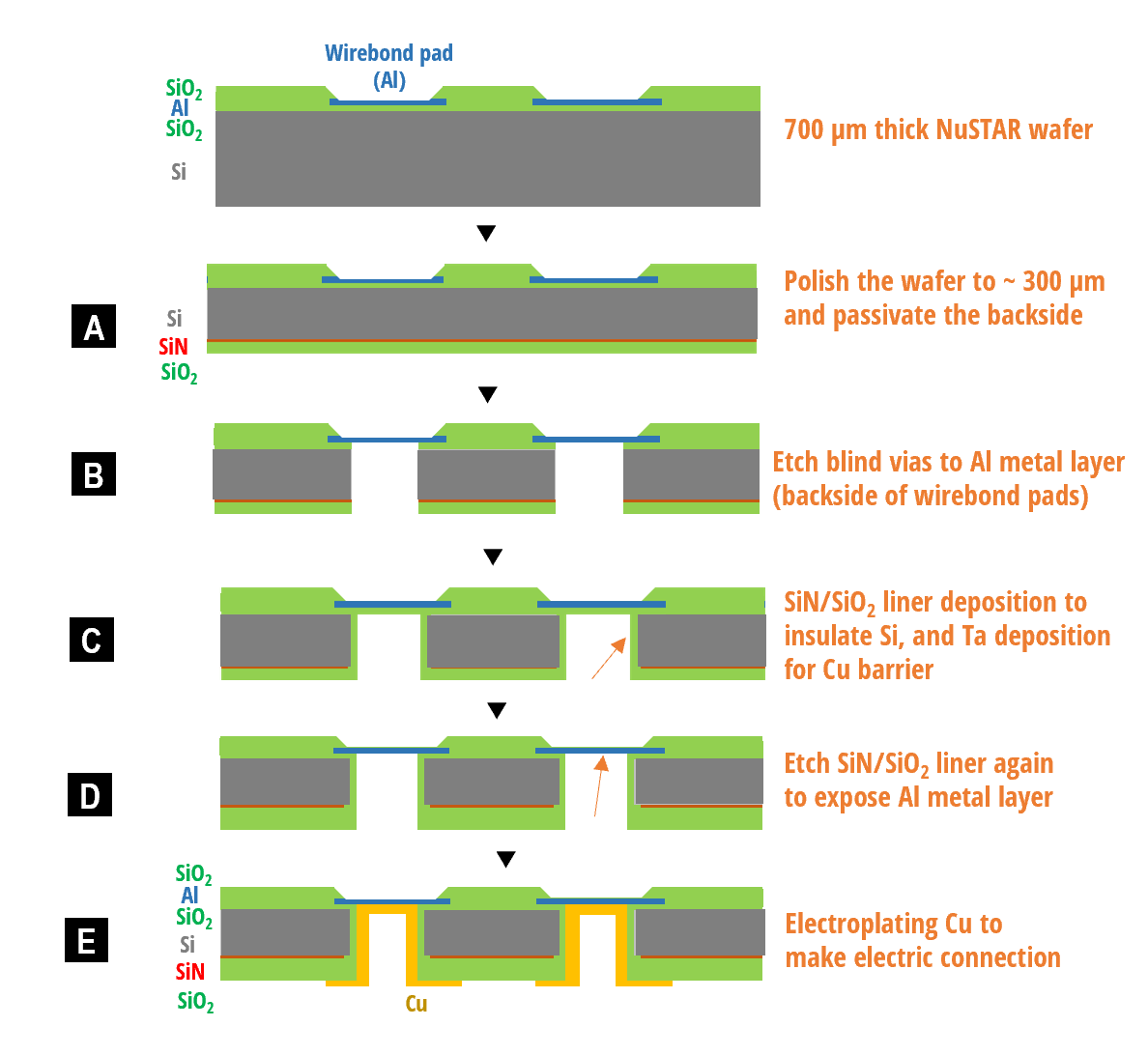}
\caption{\fixU{Back-side {\it blind}-TSV fabrication processes} by Vendor \#1: (A) Back-side polishing and passivation. (B) Blind TSV etching. (C) TSV passivation and Cu-barrier layer installation. (D) Exposing the bottom passivation of TSVs through \fixR{directionally sensitive} etching. (E) Electroplating Cu on TSVs to make electric connection.}
    \label{f:backsideTSVprocess}
\end{figure*}

Figure \ref{f:backsideTSVprocess} illustrates the \fixU{fabrication} processes of back-side {\it blind}-TSVs by Vendor \#1. To achieve a high yield, the aspect ratio of the TSV height to diameter should be kept less than $\sim$ 3 to 5. Since the wire-bond pads are about 150 \um x 120 \um, we settled on TSVs of 100 \um diameter. (A) To keep the aspect ratio low, we polish \NuSTAR wafers down to 300 \um from the  700 \um thickness \fixU{of the original NuASIC wafer}. (B) The TSV profile is chemically etched out from the back side. (C) The etched TSVs are passivated with SiO$_2$ to insulate the surrounding Si in the substrate, and a Ta barrier layer is deposited to prevent Cu diffusion into the Si substrate. (D) Apply  \fixR{directionally-sensitive Reactive Ion Etch (RIE)} to re-open the passivation deposited on the back of the wire-bond pads. (E) The TSVs are metalized with Cu through multiple stages of sputtering. The back-side traces and flip-chip pads are plated in parallel.  \fixU{A} final passivation is also applied to protect the TSVs and traces (not shown in the figure).

Initially the vertical (cylinder shape) profile was adopted for TSVs, but it turns out that with the vertical profile, \fixU{the} Cu plating process results in excessive accumulation of Cu around the entrance of the TSVs, leaving little or no Cu plating near the back of the wire-bond pads. After exploring a few options to mitigate the issue, a tapered profile (funnel shape) for TSVs was adopted, which dramatically improved the uniformity in Cu metalization. Figure~\ref{f:SEMbacksideTSV} compares the SEM images of the TSVs with vertical (left) and tapered (right) profiles. In the former, the Cu (light grey) tends to get accumulated around the opening of the TSVs  while the Cu plating lacks near the back of the wire-bond pads, resulting in poor conductivity. In \fixU{contrast, the tapered TSV has the Cu layer } uniformly plated along the entire profile of TSVs, providing \fixU{good} electric\fixU{al} connectivity.

\section{Performance of \NuSTAR ASIC with back-side \textbf{\textit{blind}}-TSVs} \label{s:performance}
For diagnostic purpose, each NuASIC is equipped with internal pulser injection capability, which can trigger the readout sequence in the ASIC. The \fixU{pulser} charges deposited in a group of user-defined pixels by the internal pulser can be readout just like normal X-ray events.  Since the pulser source is external, the overall spectral resolution of the pulser measurement is subject to stability of the external source, the build quality of the probe card used for testing ASIC dies, and the ASIC functionality. We have developed a probe card with a proper probe pin arrangement suitable for operation from the back side of ASICs through the TSVs. The card circuitry is identical to the one used for \fixU{testing the original NuASIC wafers} from the front side through the wire-bond pads.

\begin{figure}[tbh!]
   \centering
\includegraphics[width=3.0in]{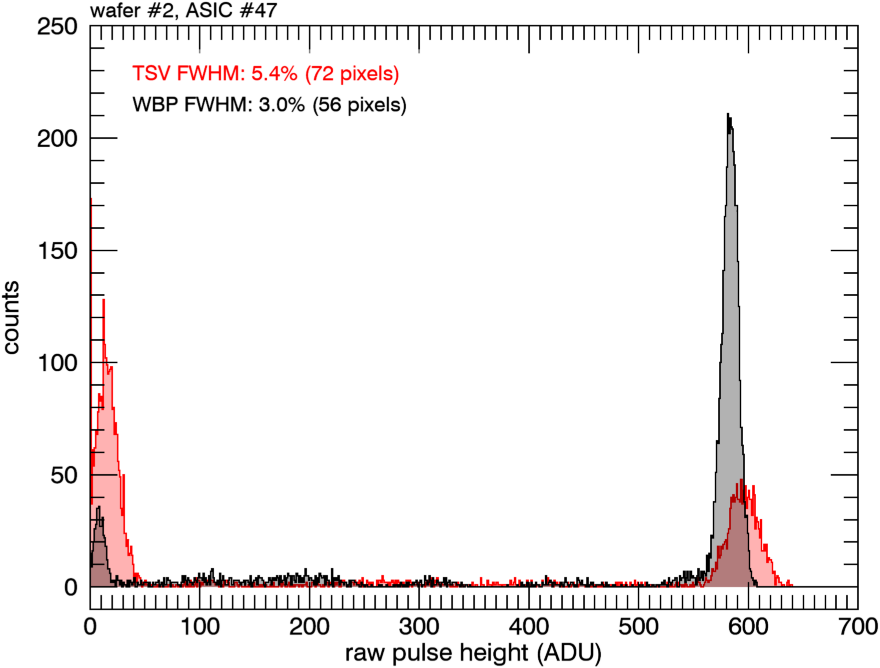}
\caption{Pulser spectra comparison between wire-bond pads and {\it blind}-TSVs with vertical profiles from Wafer \#2 Die \#47 by Vendor \#1. The measurements through the TSVs degrade the spectral resolution, and both measurements show additional noise events near 0 ADU.}
    \label{f:spectra_vertical}
\end{figure}

\begin{figure*}[tbh!]
   \centering
   \includegraphics[width=6.5in]{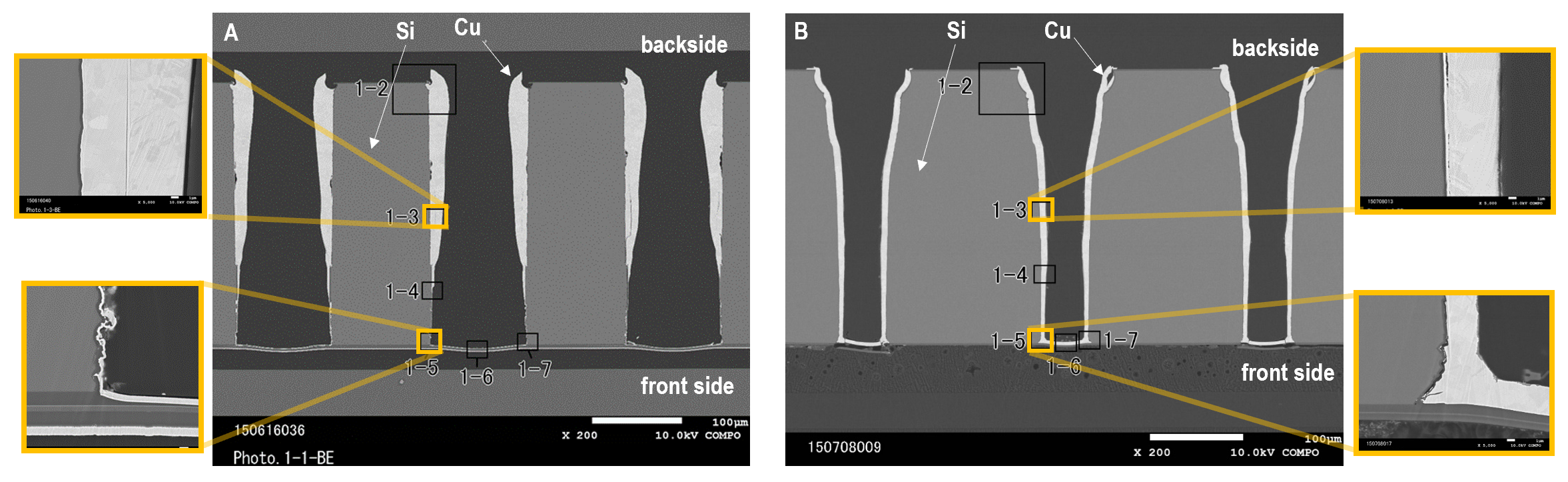}
\caption{SEM images of back-side {\it blind}-TSV profiles by Vendor \#1: (A) Vertical profile with cylinder shape. (B) Tapered profile with funnel shape.}
    \label{f:SEMbacksideTSV}
\end{figure*}

Figure~\ref{f:spectra_vertical} compares a histogram of internal pulser signals measured through the front-side wire-bond pads (black) with the same through the back-side TSVs (red) from a NuASIC (Die \#47) with {\it blind}-TSVs of vertical profile. The histograms are the sum of the pulser readout from the triggered pixels, where the first 53 pixels were injected with pulser signals. The spectral resolution of the internal pulser signals is substantially degraded when measured through TSVs (5.4\% FWHM) relative to the wire-bond pads (3\% FWHM). In addition, both measurements, either through TSVs or wire-bond pads, show noticeable noise events near 0 ADU. Many pixels also exhibit noise signals above the threshold even without pulser injection: e.g., the measurements through the TSVs triggered 72 pixels even though only 53 pixels were injected with pulser signals.

This NuASIC is from the first attempt to implement TSVs from the back side, using the vertical profile. Typically NuASICs (without TSVs) have about 90\% yield, but after the TSV insertion,  only about 5 - 6 NuASICs out of 98 NuASICs show some response to external operational commands even when they are probed from the wire-bond pads. Die \#47 is the only functioning (albeit temporary; see below) device out of 98 NuASICs from the 2 wafers that had undergone \fixU{complete}  TSV insertion processes. In fact, all the partially functioning dies also exhibit a high current draw in the power line, which suggests that the TSV insertion in the first run introduced a kind of short or current leak somewhere in the system. Die \#47 also stopped functioning after a few operations.
We suspect that the current leak permanently damaged the ASIC.

As shown in the left panel in Figure~\ref{f:SEMbacksideTSV}, inspection of the high resolution SEM images of several TSV profiles revealed uneven etching surface of the TSV profiles \fixU{especially near the bottom of the TSVs}, suggesting potential electrical shorts between the Cu metal in the TSV and the Si substrate. We could not pin-point  \fixU{any actual spot of electrical shorts} from the SEM images, which is not surprising, given the limited cross-sections one can explore through a few SEM images.  Regardless,
to mitigate the issue, the passivation layer of the TSVs were doubled for the 2nd TSV run
which utilized the tapered TSV profiles instead of the vertical profiles. 

\begin{figure}[tbh!]
   \centering
\includegraphics[width=3.0in]{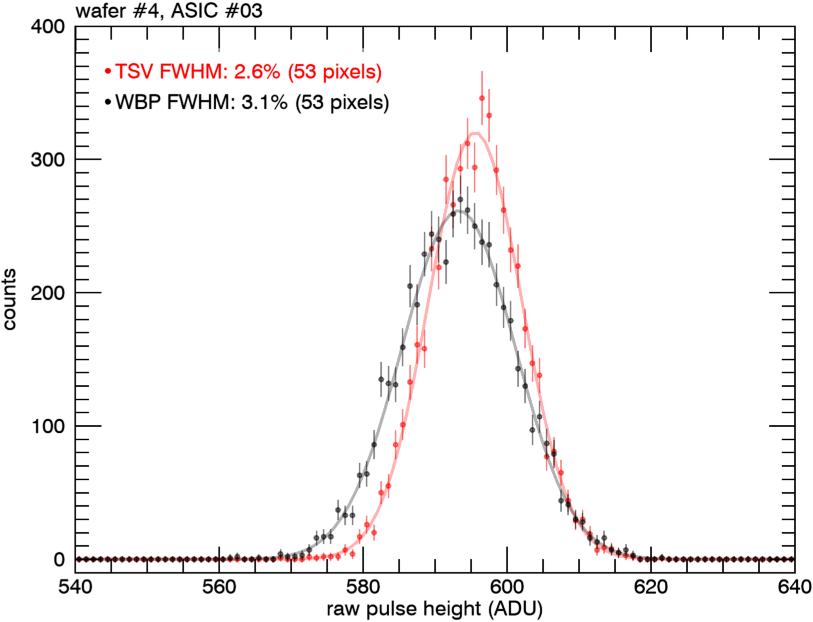}
\caption{Pulser spectra comparison between wire-bond pads and {\it blind}-TSVs with tapered profiles from Wafer \#3 Die \#3 by Vendor \#1.
The measurements through the TSVs show an improvement in the spectral resolution, compared to the measurements through the wire-bond pads.}
    \label{f:spectra_tapered}
\end{figure}

Figure~\ref{f:spectra_tapered} compares internal pulser measurements of a pixel on one of the NuASICs through wire-bond pads before TSV insertion and through TSVs after successful insertion of TSVs with tapered profiles. The spectrum measured through the TSVs shows an improvement in the spectral resolution (2.6\% FWHM) relative to the measurements conducted through the wire-bond pads before TSV insertion (3.1\% FWHM).  These ASICs with tapered profile TSVs show no excessive noise events and their performance is stable over time unlike Die \#47 in the earlier run.
Despite the fact that the signal path length is slightly longer by as much as $\sim$ 1 mm  when probed through the TSVs compared to the wire-bond pads, the improved spectral resolution in the measurements through TSVs is encouraging. In a completed detector assembly, the electrical paths through wire bonds are longer and more exposed
than those through TSVs, and thus the system with TSVs is expected to be less susceptible to external noises. 
Note that pulser measurements of NuASICs depend on the stability of the external pulser source and the build quality of the probe cards as well as the ASIC performance. Nonetheless, the results indicate that the proper TSV insertion process does not degrade the ASIC performance.

\begin{figure*}[tbh!]
   \centering
\includegraphics[width=6.0in]{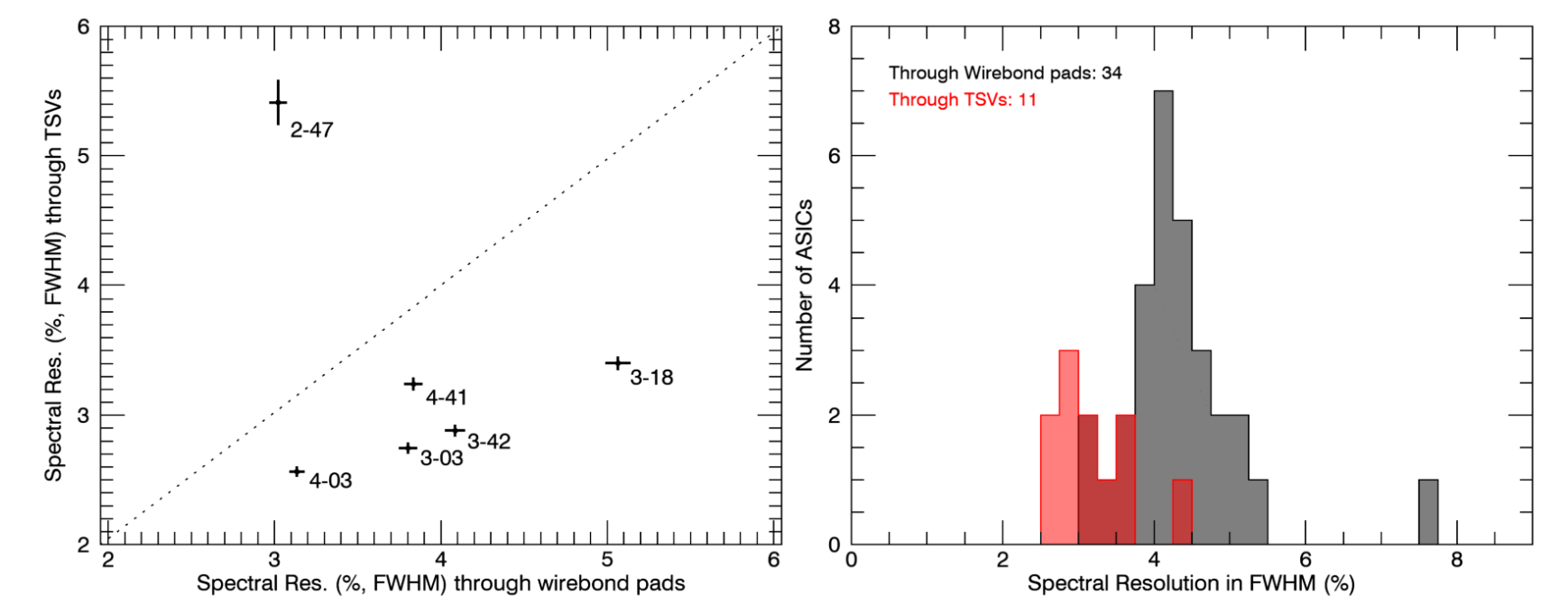}
\caption{\fixU{Spectral} resolution comparison between wire-bond pads and {\it blind}-TSVs by Vendor \#1. (Left) the ASIC dies with both measurements available. (Right) Resolution distribution of all the measurements for Wafers \#3 and \#4.}
    \label{f:resolution}
\end{figure*}

The left panel in Figure~\ref{f:resolution} compares the pulser measurements of the NuASICs where measurements through both wire-bond pads and TSVs are available for the same ASIC. Each data point is labeled by wafer and ASIC die IDs. Note the measurements through wire-bond pads for ASICs in Wafers \#3 and \#4 were conducted before TSV insertion.  TSVs in Wafer \#2 have vertical profiles, and ones in Wafers \#3 and \#4 have tapered profiles. All the NuASICs with tapered profile TSVs show improvements in the pulser resolution when measured through TSVs compared to wire-bond pads. 
The right panel shows all the measurements we have conducted for ASICs in Wafers \#3 and \#4, where the measurements through TSVs statistically outperform the measurements through wire-bond pads in spectral resolution, indicating a successful demonstration of the proof of concept for the via-last process for TSVs. 
The overall yield of the successful TSV implementation in Wafers \#3 and \#4 are about 70\% after accounting for the overall non-TSV NuASIC yield (90\%).

Despite the relatively high yield in successful TSV implementation, we were not successful in assembling fully functioning CZT detectors from these TSV NuASICs.
One of the main causes was excessive plating of back-side traces and bonding pads in part due to inexperience in trace plating of Vendor \#1.  As a result, many traces were shorted to each other, making a large number of dies unusable. Over-plating also introduced unnecessary stress in the wafers, which resulted in large cracks in the wafers during the debonding process of the carrier layer. The carrier layer is temporarily attached to protect thin wafers during the TSV implementation.
The cracks further reduced the number of workable dies.
Only 17 out of 98 NuASICs from the two wafers were properly plated for the traces and pads on the back side. In the case of Wafer \#4, the carrier layer debonding was abandoned due to the risk of cracking, which renders all the ASICs unusable for subsequent detector assembly since the carrier layer blocks access to the pixel pads on the front side.

\begin{figure}[tbh!]
   \centering
\includegraphics[width=3.0in]{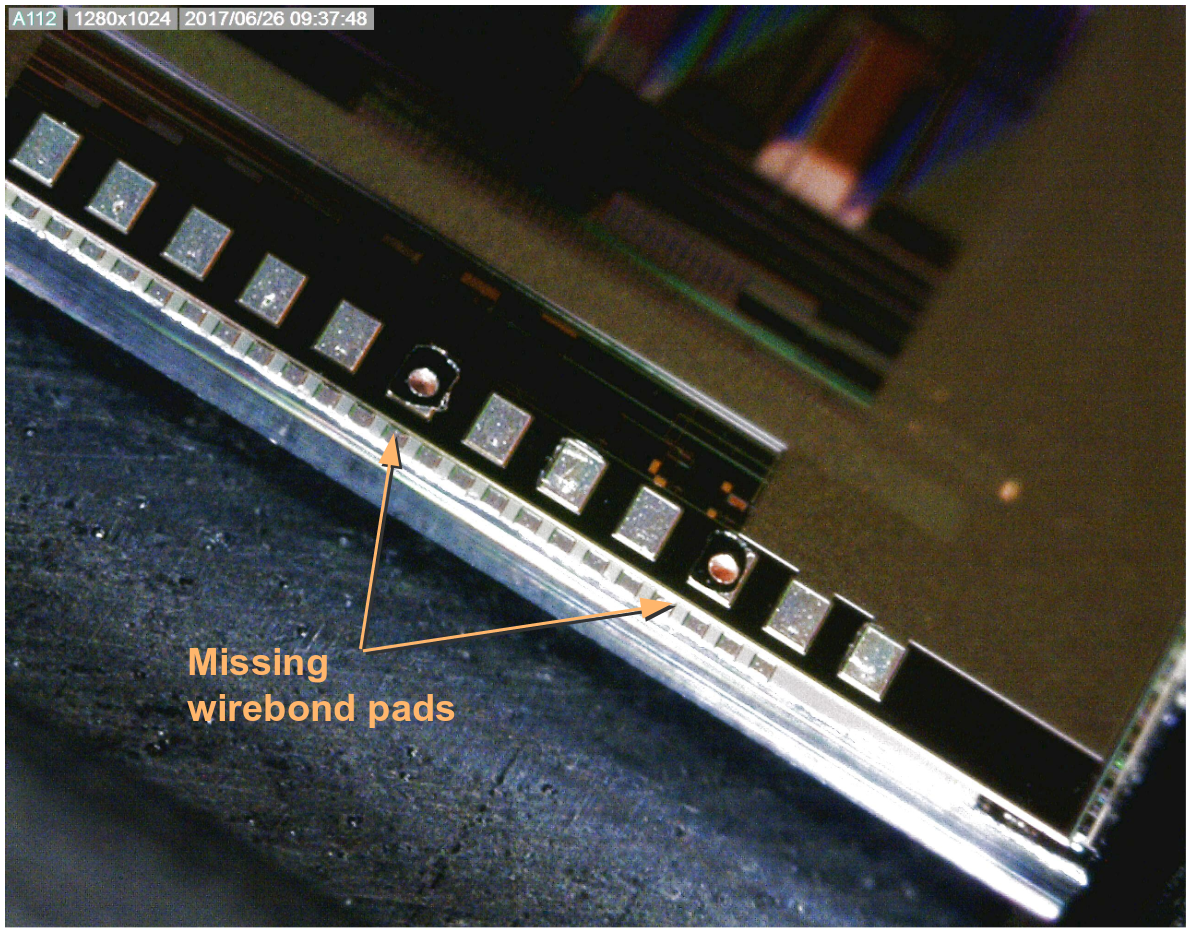}
\caption{Loss of wire-bond pads during carrier layer debonding for {\it blind}-TSV implementation by Vendor \#1. Among 12 wire-bond pads visible in this view, two pads were \fixU{delaminated} during the debonding process of the carrier layer, which revealed the TSVs underneath.}
    \label{f:wbploss}
\end{figure}

Debonding of the carrier layer from Wafer \#3 revealed another weakness in the TSV implementation from the back side. Since TSVs touch down on the back of the wire-bond pads and the majority volume of the TSVs  \fixU{is} not filled, the structural integrity of wire-bond pads \fixU{is} compromised after TSV insertion. As a result, the debonding process often \fixU{delaminates} parts or
all of some wire-bond pads (Figure~\ref{f:wbploss}), which makes the ASIC unusable either through TSVs or wire bonds. Only one (Wafer \#3 Die \#47) out of the 11 functioning TSV NuASICs survived the debonding process without loss of wire-bond pads and functions properly after the assembly with the substrate board.

Another weakness of the back-side TSV implementation is in the nature of the blind vias. 
It is difficult to reach and process the region where TSVs touch down on the wire-bond pads.  The touch-down region also tends to develop unwanted chemical reactions and leave chemical residue while undergoing the multiple processes of etching and plating. This residue is often difficult to clean. As a result, this approach requires relatively expensive \fixU{wafer processing} tools, which small vendors may not have access to. In contrast, relatively large wafer processing vendors do have these necessary tools, but they often do not show interest in small volume  development \fixU{projects} like ours. Unfortunately, 
Vendor \#1, which performed the runs described in the above sections, stopped offering TSV services needed for our development after a recent corporate restructuring, and we had to search for other vendors. We did regroup with a new vendor that retained some of key members from  Vendor \#1, who had worked on our TSV development effort.  The subsequent attempt with the new vendor to implement back-side {\it blind}-TSVs, however, was unsuccessful due to lack of necessary tools, despite a few novel ideas to address the challenges. 


\section{Front-side \textbf{\textit{through}}-TSVs and their performance} \label{s:frontside}

With the successful proof of the TSV concept, we  searched for a new approach that enables high yield in detector assembly with relatively easily accessible tools. A promising approach is the implementation of true {\it through}-\fixU{Silicon} vias from the front side of wafers (instead of back-side {\it blind}-TSVs). This approach  resolves the difficulties in the previous runs with TSV etching, metallization, and interconnection to the ASIC bond pads. We have finished the initial test run of implementing front-side TSVs on 3 NuASIC Wafers with \fixU{our current industry partner, Micross Advanced Inter-connect Technology} (AIT). 

\begin{figure*}[tbh!]
\centering
   \includegraphics[width=4.5in]{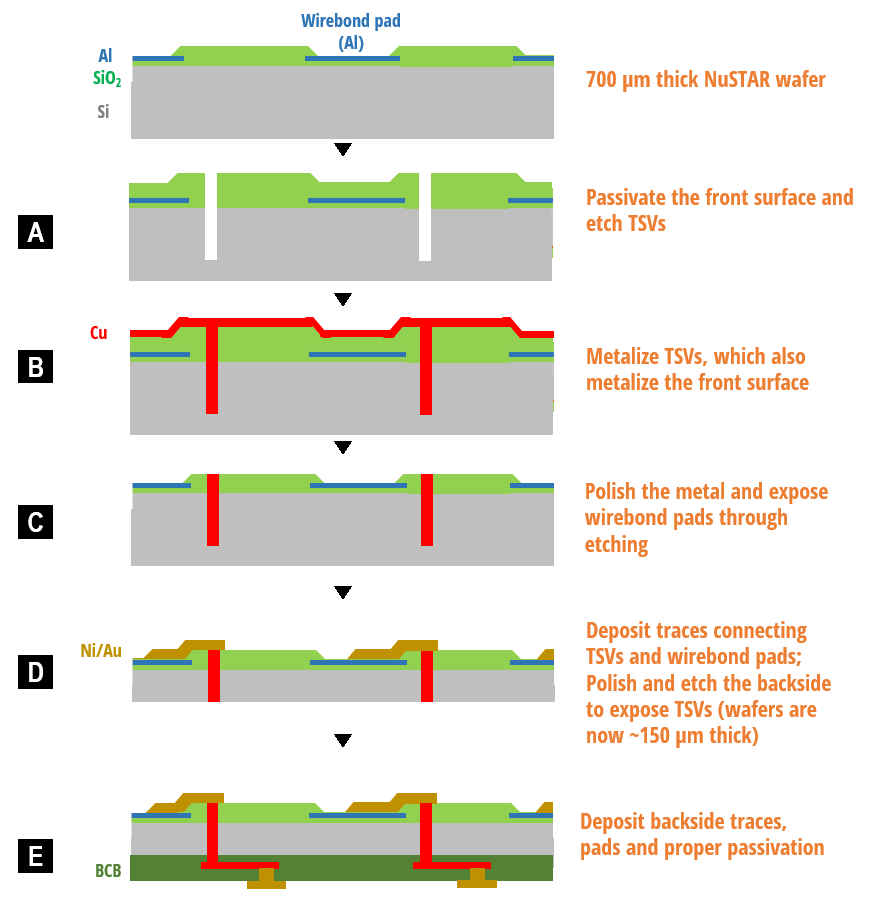}
\caption{\fixU{Front-side {\it through}-TSV fabrication processes} by Vendor \#2 (Micross AIT): (A) front-side passivation and TSV etching. (B) TSV and front-side metalization. (C) Exposing wire-bond pads through chemical-mechanical polishing, surface planarizing and etching. (D) Deposition of traces connecting TSVs and wire-bond pads, and exposing TSVs from the back side through polishing followed by etching. (E) Trace and pad deposition on the back side along with proper passivation.}
    \label{f:frontsideTSVprocess}
\end{figure*}

Figure~\ref{f:frontsideTSVprocess} illustrates the \fixU{fabrication} processes of the front-side {\it through}-TSVs. (A) The front side of the wafer is passivated and the (vertical) profiles of TSVs are etch\fixU{ed} out {\it next} to wire-bond pads. (B) After properly passivating the TSVs with SiO$_2$ and depositing the Cu-barrier layers (TiN) \fixU{to prevent Cu diffusion into the Si wafer, the} TSVs are Cu-filled along with the front surface. (C) \fixU{The} front-side \fixU{Cu} deposit is polished off so that the wafer surface is planarized with \fixU{SiO$_2$} remaining over the Al pad, and then, both wire-bond \fixU{pads} and pixel pads are exposed through etching. (D) The Ni/Au traces connecting TSVs to wire-bond pads are deposited on the front side, and then the back side is polished and etched to expose the TSVs. (E) Passivate the back side and deposit the proper metal traces and \fixU{solder bump} pads on the back side for flip-chip bond. 

This approach does not disturb the Si substrate below the wire-bond pads on the front side of the original NuASIC, so there is no risk of weakening or losing any wire-bond pads in the process. The Cu metal  is deposited inside the TSVs 
\fixU{after} the TSVs profiles are etched {\it blind} (C), but the blind end of the TSVs are revealed again through
dry etch after polishing, so there is no concern of developing any complex chemical residues that require careful cleaning and monitoring. This in turn enables use of relatively small vias (20 $\mu$m diameter). These small TSVs are fully Cu-filled, improving the mechanical integrity of each die. 

\begin{figure}[tbh!]
   \centering
\includegraphics[width=6.0in]{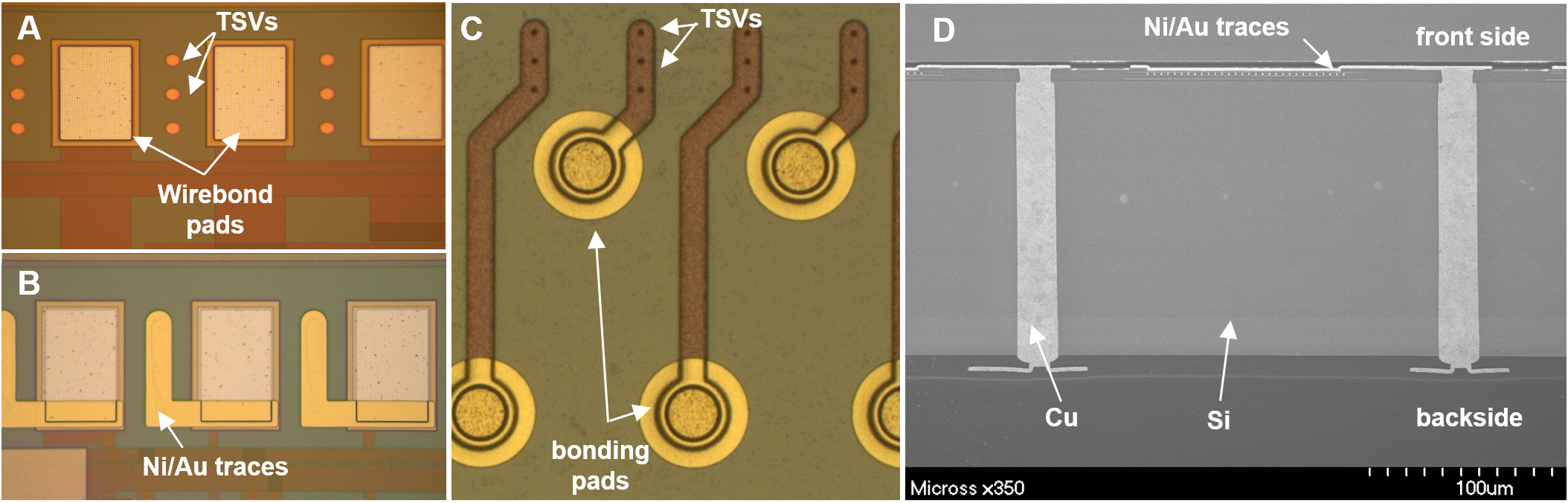}
\caption{Progress of the {\it through}-TSV implementation by Vendor \#2 (Micross AIT): (A) the front side after TSV Cu-filling and polishing, showing TSVs next to wire-bond pads (3 TSVs per wire-bond pad were employed), (B) the front side with traces connecting TSVs to wire-bond pads, (C) the back side showing TSVs, solder bump pads and connecting traces, and (D) an SEM image of the TSVs.}
\label{f:frontTSVfigure}
\end{figure}

\fixU{Our r}elatively small TSVs allow \fixU{3} TSV insertions for each wire-bond pad, improving \fixU{the} yield in conductivity. Figure~\ref{f:frontTSVfigure} shows a small region around a few wire-bond pads during TSV implementation: (A) the front side after Cu-filling TSVs (3 TSVs per wire-bond pad), (B) the front side after Ni/Au trace layout connecting TSVs to an edge of each wire-bond pad, (C) the bottom side showing the TSVs, flip-chip style pads, and connecting traces and (D) an SEM image of the TSV cross section.  \fixU{As mentioned,} we inserted 3 TSVs for each wire-bond pad in between \fixU{each wire-bond pad. We} laid out the connecting traces between TSVs and the wire-bond pads. The majority of the original Al wire-bond pads are exposed so that in principle wire bonding can be applied to these ASICs if needed. The pads for flip-chip style bonding on the back side are now made in a circular shape (instead of the \fixU{original} rectangular shape; see Figure~\ref{f:backsideTSVconcept}C) in order to improve the uniformity in the \fixU{spherical} solder ball attachment and spread, which in turn improves bonding quality. The pads are coated with Ni and \fixU{Au} to avoid oxidization.

\begin{figure}[tbh!]
   \centering
\includegraphics[width=3.0in]{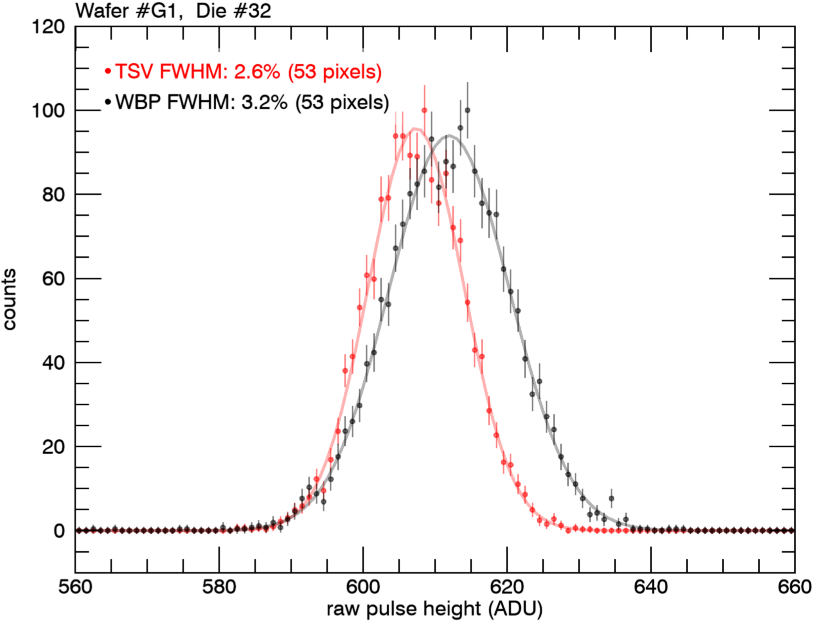}
\caption{Pulser performance of a NuASIC with front-side TSVs  by Vendor \#2 (Micross AIT): measurements through TSVs (red) in comparison with the same through wire-bond pads (black).}
    \label{f:frontsideperformance}
\end{figure}

Figure~\ref{f:frontsideperformance} shows the pulser performance measured through the TSVs (red) in comparison with the same through the wire-bond pads (black). The measurements through the wire-bond pads were done before the TSV implementation. Some of the measurements through the TSVs were done using the ASIC Test Setup (ATS) developed at Harvard \cite{Violette18,Violette20}. The ATS is designed to enable rapid test and screening of TSV ASICs through quick and easy alignment of pogo-pins and the back-side pads or solder bumps on the pads of the ASIC. 
The distribution is a sum of the pulser data from the first 53 pixels of Wafer \#G1 and Die \#32. This demonstrates
the successful operations in the ASICs with front-side {\it through}-TSVs.  The overall spectral resolution
improved again with the TSVs relative to the operation through wire-bond pads.

Among 15 font-side TSV NuASICs that were tested so far, 4 NuASICs show a good pulser performance (similar to Figure~\ref{f:frontsideperformance}) in the initial pulser runs, showing about 30\% of yield. While this yield appears to be small relative to the earlier run, we believe that the yield is nearly 100\% in terms of the conductivity of TSVs. In this initial run for the front-side TSVs, a set of daisy chained TSVs
are inserted to each ASIC die for diagnostic purpose, where the electric\fixU{al} conductivity and isolation of TSVs can be tested \cite{Ovental20}. 
According to the conductivity test of the daisy chain, the average resistivity of TSVs is less than 1 $\Omega$ with practically 100\% yield. The cause of the relatively low yield in the overall functionality is suspected to be the insufficient isolation or insulation of the TSVs.

 In fact, all of the tested ASICs draw the increasingly higher current in the power line over time during the pulser tests, and eventually became unresponsive after about 10 minutes of operation. Subsequent power cycle would start the ASIC at a relatively higher current draw state even after a few days of non-operation, and the ASICs become unresponsive within a minute. This result is consistent with the isolation tests of the daisy chain TSVs, where the current leak between relatively nearby TSVs (electrically unconnected) was observed to increase over time \cite{Ovental20}.

While we are investigating the root cause of this current instability, the observed symptom\fixU{s} resemble what we had experienced with the very first run of the back-side {\it blind}-TSV implementation, where all the dies draw high currents in the power line and their functionality was unstable.  The most likely cause is the insufficient thickness for \fixR{the SiO$_2$} insulation and/or Cu barrier layer \fixU{surrounding each TSV as well as between the Cu traces and underlying Si wafer}. In fact, the first run of the back-side {\it blind}-TSV implementation had issues in both TSV conductivity and isolation, which resulted in a poor spectral resolution (Figure~\ref{f:spectra_vertical}) and a low yield ($\sim$ 1\%).  It is clear that the front-side {\it through}-TSVs in this first trial run have good conductivity but poor insulation according to both the pulser and daisy chain tests.  We plan to increase the thickness of \fixR{SiO$_2$} insulation and Cu barrier layers, which had resolved the similar issue for the back-side {\it blind}-TSVs.   As we learn more about the cause of the high \fixU{current} draw and instability, we will further revise the processes, which will be applied in the next run for the front-side {\it through}-TSV implementation. Given the good conductivity of the front-side TSVs, we expect that the overall yield for successful NuASICs with front-side TSVs will exceed the 70\% yield in the back-side TSV NuASICs of the 2nd-run once we resolve the TSV isolation issue.



\section{Summary and Detector Assembly with TSVs} \label{s:future}

We presented the \fixU{first} successful proof of concept for TSVs \fixU{to replace wire bonds for readout and control of ASICs bonded to CZT imaging detectors.   We used} two approaches. While both  are viable, the approach using front-side {\it through}-TSVs resolves major weaknesses in the other approach using back-side {\it blind}-TSVs, enabling an easier \fixU{and cleaner} path for \fixU{both TSV insertion and final} detector assembly.

\begin{figure*}[bh!]
   \includegraphics[width=6.4in]{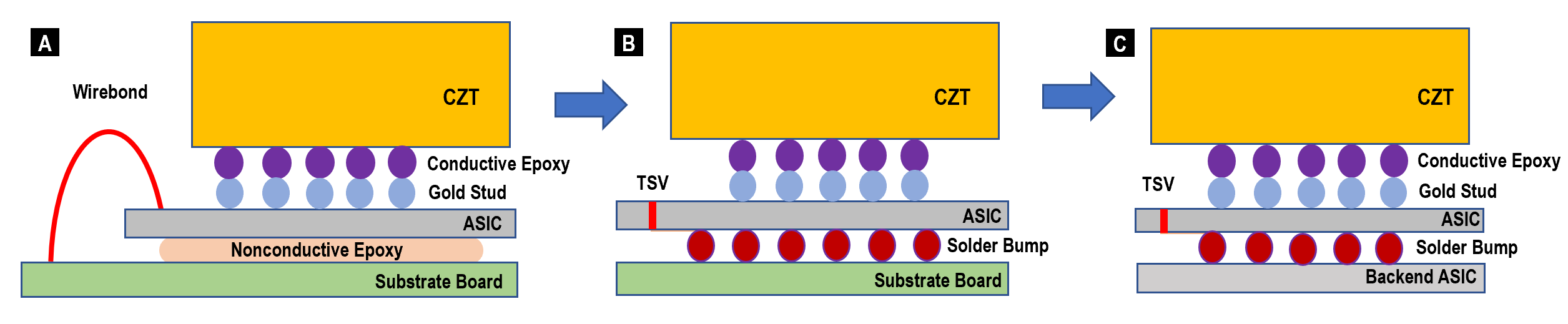}
\caption{(A) CZT and ASIC assembly using wire bonds. (B) The same with TSVs. (C) CZT, ASIC and backend ASIC assembly using 3-D stacking.}
    \label{f:CZTdetectordevelop}
\end{figure*}

With the TSV NuASICs, we can envision true 3-D stacking of the backend electronics for CZT detector assembly in future as illustrated in Figure~\ref{f:CZTdetectordevelop}. The current design of our CZT detector assembly employs a set of Field Programmable Gate Arrays (FPGAs) and complex programmable logic devices (CPLDs)  to
control and readout the NuASICs. These logic devices are mounted in the substrate board where the NuASICs are mounted and connected through wire bonds (Figure~\ref{f:CZTdetectordevelop}A). With TSVs, assembly between \fixU{the CZT crystal, the NuASIC and the substrate board} is  achieved essentially through a series of flip-chip style bonding (Figure~\ref{f:CZTdetectordevelop}B). Once the operational code of the logic devices are well defined and frozen, the front end of the FPGAs can be turned into another signal-processing ASIC, which can be directly 3-D stacked underneath the ASIC using TSVs (Figure~\ref{f:CZTdetectordevelop}C). The design of NuASIC itself can be further optimized to minimize the gap between the CZT crystals. One can go a step further, the layout of both the NuASIC and the signal processing ASIC can be designed with TSVs in mind. 
The TSV technology enables easy, robust and low cost assembly and packaging concept to a wide range of detector and sensor systems. 


\section{Acknowledgement}

This work was supported by NASA APRA grant NNX17AE62G. 



\bibliography{references} 
\bibliographystyle{spiejour}


\section*{Biographies}

{\bf Jaesub Hong} is a Research Associate at Harvard University. He has nearly 20 years of experience in development of X-ray telescopes for high energy astrophysics and planetary science.  His current focus is  the development of advanced hard X-ray detectors for next generation wide-field hard X-ray telescopes for time domain astrophysics and the miniature lightweight X-ray optics for planetary science.  He received a Ph.D.~degree in Physics from Columbia University. He has (co)authored over 40 publications. 

{\bf Jonathan (Josh) Grindlay} is the Robert Treat Paine Professor of Astronomy at Harvard. He received his BA in Physics from Dartmouth (1966) and PhD in Astrophysics from Harvard  (1971). He joined the Faculty in 1976 and  Chaired the Department in 1985-91 and 2001-03. His primary interest is black hole time variability, accretion physics, accreting black hole (both stellar and supermassive) populations and formation as measured with wide-field coded aperture imaging X-ray telescopes (ultimately full-sky) and optical/IR imaging/spectroscopy. He has over 434 refereed Journal papers.

{\bf Branden Allen} received his Ph.D.~degree in physics from U.C. Irvine in 2007 and is currently a Research Associate at Harvard University with over 20 years of experience in the development, deployment and operation of ground- and space-based telescopes for high-energy X/$\gamma$-ray astronomy and planetary science.  His current research is focused on the development and deployment of next generation detector systems and telescopes to probe high energy astrophysical phenomena and for future planetary exploration.  

{\bf Daniel Violette} is a graduate student at Harvard University working with the High-Resolution Energetic X-ray Imager (HREXI) team, with interests in high-energy time domain astrophysics and instrumentation. Daniel is supported by the Future Investigators in NASA Earth and Space Science and Technology (FINESST) Fellowship to further develop HREXI detector sensitivity at low energy thresholds.

{\bf Hiromasa Miyasaka} is a staff scientist at California Institute of Technology. He received a Ph.D.~in Physics (2000) from Saitama University in Japan. He has over 20 years of experience in development of particles and X-ray detectors for the cosmic ray and high-energy astrophysics. Since 2006, his work has focused on CdZnTe and CdTe detectors and readout ASIC development. He is one of the primary detector scientists for the \NuSTAR mission.

{\bf Dean Malta} is a Program Manager at Micross AIT. His work has spanned nearly 30 years in microelectronics packaging, fiber optic components, and wide bandgap semiconductor devices. His current focus is the development of 3D and heterogeneous integration technologies for advanced electronic modules and photonic sensors. Dean received a B.S. in Electrical Engineering from Wilkes University. He is a Senior Member of IEEE, holds 6 U.S. patents, and has (co)authored over 75 publications.

{\bf Jennifer Ovental} is a Development Engineer at Micross AIT. Her primary focus is on 3D and Heterogeneous Integration, particularly in the areas of through-silicon via technology and microbump processes for ultra-fine pitch chip bonding. She received her BS in Chemical Engineering (2012) from Lafayette College and her Ph.D.~in Chemical Engineering (2018) from North Carolina State University. Her graduate research and published work was predominately focused on atomic layer deposition.

{\bf David Bordelon} is a Development Engineer in Advanced Packaging and 3D Integration for government and commercial programs with Micross AIT. He received his BS in Applied Science (1994) and Ph.D.~(2009) in Materials Science from UNC-Chapel Hill. His published articles include Ph.D.~work on biomedical applications of carbon nanotube field emission and post-doctoral research in nanomaterial-mediated cancer thermotherapy at Johns Hopkins. He gained extensive microfabrication expertise during 10+ years of prior Optical-MEMS \& Microsystems development.

{\bf Daniel Richter} is a Development Engineer at Micross AIT and is responsible for the development and implementation of advanced assembly and semiconductor packaging technology. He has extensive design experience and takes a lead role in photomask layout. He holds multiple patents relating to semiconductor packaging technology. Daniel received a B.S. degree in Material Science Engineering and M.S. degree in Nanoengineering from North Carolina State University.

\section*{Tables}

N/A

\section*{Figures}

\end{document}